 \documentclass[12pt]{article}
 \usepackage[cp1251]{inputenc} 
 \usepackage[english, russian]{babel} 
 \usepackage{amsmath, latexsym, amssymb, bm, array, graphics, eucal,
 amsfonts,}
 \makeatletter

\renewcommand{\Im}{\mathop{\rm Im\,}}
\makeatother

\usepackage[dvips]{graphicx}

\begin{document}
\thispagestyle{empty}
\large
\renewcommand{\abstractname}{Abstract}
\renewcommand{\refname}{\begin{center}
 REFERENCES\end{center}}

\makeatother

\begin{center}
\bf The kinetic one-dimensional equation with frequency of collisions,
affine depending on the module molecular velocity
\end{center} \medskip

\begin{center}
  \bf
  A. L. Bugrimov\footnote{$fakul-fm@mgou.ru$},
  A. V. Latyshev\footnote{$avlatyshev@mail.ru$} and
  A. A. Yushkanov\footnote{$yushkanov@inbox.ru$}
\end{center}\medskip

\begin{center}
{\it Faculty of Physics and Mathematics,\\ Moscow State Regional
University, 105005,\\ Moscow, Radio str., 10A}
\end{center}\medskip

\begin{abstract}
The one-dimensional kinetic equation  with
integral of collisions type BGK (Bhatnagar, Gross and Krook)
and frequency of collisions affine depending on the module
of molecular velocity is constructed.

Laws of preservation of number of particles,
momentum and energy at construction equation are used.

Separation of variables leads to the characteristic equation.
The system of the dispersion equations is entered. Its determinant
is called as dispersion function. It is investigated continuous and
discrete spectra of the characteristic equation.

The set of zero of the dispersion equation makes the discrete
spectrum of the characteristic equation. The eigen
solutions of the kinetic equation corresponding
to discrete spectrum are found.

The solution of the characteristic equation in space of the generalized
functions leads to eigen functions corresponding to the continuous
spectrum.

Results of the spent analysis  in the form of the theorem about
structure of the general solution of the entered kinetic equation
are formulated.


\medskip


\end{abstract}

\begin{center}
\bf  Introduction. Statement of problem and basic equations
\end{center}

Analytical solutions of whole
some boundary problems (temperature and density jumps,
various slidings, et cetera) the kinetic theory of gas with
use of BGK equation with constant
frequency of collisions are received to the present time
\cite{1}-\cite{5}.

Approximation of a constancy of frequency of collisions far not
always is possible to consider as adequate to a problem.
In this connection attempts become
to consider more the general, than BGK, models. In particular,
the problem about isothermal sliding is considered for
enough a wide class of BGK-models \cite{1}.

The case with the frequency of collisions proportional to the
module of molecular velocity
(i.e. with constant length of the free
path) is considered also. In this approach problems about jumps
of temperature and concentration  \cite{6}, and as more the general,
than BGK, models (see, for example, \cite{7}--\cite{9})
are considered.

At the same time there is an unresolved problem about temperature jump
and concentration with use of the BGK-equation with
any dependence of frequency on velocity, in spite
on obvious importance of the solution of a problem in similar statement.

In the present work attempt to promote in this
direction  becomes. Here the case of the affine
dependence of frequency of collisions on module velocity of molecules in
models of one-dimensional gas is considered.
Model of one-dimensional gas widely
it was used in a number of works \cite{10}--\cite{12} for gave the good
the consent with experiment.

Let us begin with the general statement. Let gas occupies
half-space $x> 0$. The surface temperature $T_s $  and
concentration of sated steam of a surface $n_s $ are set.
Far from a surface gas moves with some velocity $u $,
being velocity of evaporation (or condensation),
also has the temperature gradient
$$
g_T=\Big(\dfrac{d\ln T}{dx}\Big)_{x=+\infty}.
$$

It is necessary to define jumps of
temperature and concentration depending on velocity and
temperature gradient.

In a problem about weak evaporation
it is required to define temperature and concentration jumps
depending on velocity, including a temperature gradient
equal to zero, and velocity of evaporation (condensation) is enough small.
The last means, that
$$
u \ll v_T.
$$

Here $v_T$ is the heat velocity of molecules, having order
of sound velocity order,
$$
v_T=\dfrac{1}{\sqrt{\beta_s}}, \qquad \beta_s=\dfrac{m}{2k_BT_s},
$$
$m$ is the mass of molecule, $k_B$ is the Boltzmann constant.

In a problem about temperature jump it is required to define
temperature and concentration jumps
depending on a temperature gradient, thus
evaporation (condensation) velocity
it is considered equal to zero, and the temperature gradient is
considered  as small.
It means, that
$$
lg_T\ll 1, \qquad l=\tau v_T,\qquad \tau=\dfrac{1}{\nu_0},
$$
where $l$ is the mean free path of gas molecules,
$\tau$ is the mean relaxation time, i.e. time between two
consecutive collisions of molecules.

Let us unite both problems  (about weak evaporation (condensation) and
temperature jump) in one. We will assume that the gradient
of temperature is small (i.e. relative difference
of temperature on length of mean free path is small) and the
velocity of gas in comparison with sound velocity is small. In
this case the problem supposes linearization and  distribution function
it is possible to search in the form
$$
f(x,v)=f_0(v)(1+h(x,v)),
$$
where
$$
f_0(v)=n_s\Big(\dfrac{m}{2\pi k_BT_s}\Big)^{1/2}
\exp \Big[-\dfrac{mv^2}{2k_BT_s}\Big]
$$
is the absolute Maxwellian.

We take the linear kinetic equation which has been written down rather
functions $h (x, v) $, with integral of collisions of
relaxation type, in \cite{1} integral of collisions BGK named also
(Bhatnagar, Gross and Krook), and having the following form
$$
v\dfrac{\partial h}{\partial x}=\nu(v)\Big[l_0[v]+
2\dfrac{v}{v_T}l_1[v]+\Big(\dfrac{v^2}{v_T^2}-\beta\Big)l_2[v]-
h(x,v)\Big].
\eqno{(1.1)}
$$

Here
$l_\alpha[h]$ ($\alpha=0,1,2$) is the any constants,
subject to definition from
laws of preservation of number of particles (numerical density),
an momentum and energy, $ \nu(v) $ is the collision frequency affine
depending on module molecular velocity,
$$
\nu(v)=\nu_0\Big(1+\sqrt{\pi}a\sqrt{\dfrac{m}{2k_BT_s}}|v|\Big),
$$
$a$ is the positive parameter, $0\leqslant a<+\infty$.

The right part of the equation (1.1) is the linear integral
of collisions, spread out on collision invariants
$$
\hspace{-2.1cm}\psi_0(v)=1,
$$
$$
\psi_1(v)=2\sqrt{\dfrac{m}{2k_BT_s}}v,
$$
$$
\psi_2(v)=\dfrac{mv^2}{2k_BT_s}-\beta.
$$

The constant $ \beta $ is finding from an orthogonality condition
of invariants $ \psi_0(v) $ and $ \psi_2(v) $. Orthogonality here
it is understood as equality to zero of scalar product with weight
$\rho(v)=\nu(v)\exp\Big(-\dfrac{mv^2}{2k_BT_s}\Big)$
$$
(f,g)=\int\limits_{-\infty}^{\infty}\nu(v)
\exp\Big(-\dfrac{mv^2}{2k_BT_s}\Big)
f(v)g(v)dv.
$$

Let us pass in the equation (1.1) to dimensionless velocity
$$
C=\sqrt{\beta}v=\dfrac{v}{v_T}
$$
and dimensionless coordinate
$$
x'=\nu_0 \sqrt{\dfrac{m}{2k_BT_s}}x=\dfrac{x}{l}
$$

The variable $x'$ let us designate again through $x$.

In the dimensionless variables we will rewrite the equation
(1.1) in the form
$$
C\dfrac{\partial h}{\partial x}=(1+\sqrt{\pi}a|C|)\Big[l_0[h]+
2Cl_1[h]+(C^2-\beta)l_2[h]-h(x,C)\Big].
\eqno{(1.2)}
$$

The constant $\beta $ is definded, how it was already specified,
from the condition
$$
(\psi_0,\psi_2)=\nu_0
\int\limits_{-\infty}^{\infty}e^{-C^2}(1+\sqrt{\pi}a|C|)(C^2-\beta)dC=0.
$$

From here we receive, that
$$
\beta=\beta(a)=\dfrac{2a+1}{2(a+1)}.
$$

\begin{center}
\bf 2. Laws of preservation and transformation of the kinetic
equation
\end{center}

The modelling integral of collisions should satisfy to laws
preservations of number of particles (numerical density), momentum and
energy
$$
(\psi_\alpha,M[h])\equiv \nu_0\int\limits_{-\infty}^{\infty}
e^{-C^2}(1+\sqrt{\pi}a|C|)M[h]\psi_\alpha(C)dC=0,
\eqno{(2.1)}
$$
where $\alpha=0,1,2$, $\rho(C)M[h]$ is the model collision
integral,
$$
M[h]=l_0[h]+
2Cl_1[h]+(C^2-\beta)l_2[h]-h(x,C).
$$

From the first equation from (2.1), i.e. preservation law of
number of particles
$(\psi_0,M[h])=0$ we receive that
$$
l_0[h]=\dfrac{(1,h)}{(1,1)}.
$$
Here
$$
(1,1)=\nu_0\int\limits_{-\infty}^{\infty}e^{-C^2}(1+\sqrt{\pi}a|C|)dc=
\nu_0 \sqrt{\pi}(a+1),
$$
$$
(1,h)=\nu_0\int\limits_{-\infty}^{\infty}e^{-C^2}
(1+\sqrt{\pi}a|C|)h(x,C)dC.
$$

It means, that
$$
l_0[h]=\dfrac{1}{\sqrt{\pi}(a+1)}\int\limits_{-\infty}^{\infty}e^{-C^2}
(1+\sqrt{\pi}a|C|)h(x,C)dC.
$$

From second equation from (2.1), i.e. preservation law of momentum
$(\psi_1,M[h])=0$ we receive that
$$
2l_2[h]=\dfrac{(C,h)}{(C,C)},
$$
where
$$
(C,C)=\nu_0\int\limits_{-\infty}^{\infty}e^{-C^2}
(1+\sqrt{\pi}a|C|)C^2dC=\nu_0\dfrac{\sqrt{\pi}}{2}(2a+1),
$$
$$
(C,h)=\nu_0\int\limits_{-\infty}^{\infty}e^{-C^2}
(1+\sqrt{\pi}a|C|)Ch(x,C)dC.
$$

Therefore,
$$
2l_1[h]=\dfrac{2}{\sqrt{\pi}(2a+1)}\int\limits_{-\infty}^{\infty}e^{-C^2}
(1+\sqrt{\pi}a|C|)Ch(x,C)dC.
$$

From third equation from (2.1), i.e. preservation law of energy
$(\psi_2,M[h])=0$ we receive that
$$
(\psi_2, M[h])=\nu_0\int\limits_{-\infty}^{\infty}e^{-C^2}
(1+\sqrt{\pi}a|C|)(C^2-\beta)^2dC-$$$$-
\nu_0\int\limits_{-\infty}^{\infty}e^{-C^2}
(1+\sqrt{\pi}a|C|)(C^2-\beta)h(x,C)dC=0,
$$
whence
$$
l_2[h]=\dfrac{(C^2-\beta,h)}{(C^2-\beta,C^2-\beta)}.
$$

Here
$$
(C^2-\beta,C^2-\beta)=\nu_0\int\limits_{-\infty}^{\infty}e^{-C^2}
(1+\sqrt{\pi}a|C|)(C^2-\beta)^2dC=$$$$=
\nu_0\sqrt{\pi}\dfrac{4a^2+7a+2}{4(a+1)},
$$
$$
(C^2-\beta,h)=\nu_0\int\limits_{-\infty}^{\infty}e^{-C^2}
(1+\sqrt{\pi}a|C|)(C^2-\beta)h(x,C)dC.
$$
From last equalities it is had
$$
l_2[h]=\dfrac{4(a+1)}{\sqrt{\pi}(4a^2+7a+2)}
\int\limits_{-\infty}^{\infty}e^{-C^2}
(1+\sqrt{\pi}a|C|)(C^2-\beta)h(x,C)dC.
$$

Let us return to the equation (1.2) and by means of received above equalities
let us transform this equation to the form
$$
C\dfrac{\partial h}{\partial x}+(1+\sqrt{\pi}a|C|)h(x,C)=
$$
$$
=(1+\sqrt{\pi}a|C|)\dfrac{1}{\sqrt{\pi}}\int\limits_{-\infty}^{\infty}
e^{-C'^2}(1+\sqrt{\pi}a|C'|)q(C,C',a)h(x,C')dC'.
\eqno{(2.2)}
$$

Here $q(C,C',a)$ is the kernel of equation,
$$
q(C,C',a)=r_0(a)+r_1(a)CC'+r_2(a)(C^2-\beta(a))(C'^2-\beta(a)),
$$
$$
r_0(a)=\dfrac{1}{a+1},\qquad r_1(a)=\dfrac{2}{2a+1},\qquad
r_2(a)=\dfrac{4(a+1)}{4a^2+7a+2}.
$$

{\sc Corollary.}
Let us notice, that at $a\to 0$ the equation (2.2)
passes in the equation

$$
C\dfrac{\partial h}{\partial x}+h(x,C)=\dfrac{1}{\sqrt{\pi}}
\int\limits_{-\infty}^{\infty}e^{-C'^2}q(C,C',0)h(x,C)dC
$$
with kernel
$$
q(C,C',0)=1+2CC'+2\Big(C^2-\dfrac{1}{2}\Big)\Big(C'^2-\dfrac{1}{2}\Big).
$$

This equation is one-dimensional BGK-equation with  constant
frequency of collisions.

Let us consider the second limiting case of the equation (2.2).
We will return to  expression of frequency of collisions
also we will copy it in the form
$$
\nu(C)=\nu_0(1+\sqrt{\pi}a|C|)=\nu_0+\nu_1|C|,
$$
where
$$
\nu_1=\sqrt{\pi}\nu_0 a.
$$

Let us tend $ \nu_0$ to zero. In this limit the quantity $a $ tends to
$ + \infty $, because
$$
a=\dfrac{\nu_1}{\sqrt{\pi}\nu_0}.
$$

It is easy to see, that in this limit
$$
\lim\limits_{a\to+\infty}(1+\sqrt{\pi}a|C'|)q(C,C',a)=
\sqrt{\pi}|C'|q_1(C,C'),
$$
where
$$
q_1(C,C')=1+CC'+(C^2-1)(C'^2-1).
$$

The equation (2.2) will thus be copied in the form
$$
\dfrac{C}{|C|}\dfrac{\partial h}{\partial x_1}+h(x_1,C)=
\int\limits_{-\infty}^{\infty}e^{-C'^2}|C'|q_1(C,C')dC'.
$$
In this equation
$$
x_1=\nu_1\sqrt{\beta_s}x=\dfrac{x}{l_1},\qquad l_1=v_T\tau_1,\qquad
\tau_1=\dfrac{1}{\nu_1}.
$$

This equation is the one-dimensional kinetic equation with
the frequency of collisions proportional to the module of the molecular
velocity.

In the equation (2.2) we will carry out variable replacement
$\sqrt{\pi}a\to a$ and transform  the received equation in the form
$$
\dfrac{C}{1+a|C|}\dfrac{\partial h}{\partial x}+h(x,C)=$$$$=
\int\limits_{-\infty}^{\infty}e^{-C'^2}(1+a|C'|)q(C,C',a)h(x,C')dC'.
\eqno{(2.3)}
$$

In this equation $q(C,C',a)$ is the kernel of equation,
$$
q(C,C',a)=r_0(a)+r_1(a)CC'+r_2(a)(C^2-\beta(a))(C'^2-\beta(a)),
$$
where
$$
r_0(a)=\dfrac{1}{a+\sqrt{\pi}},\quad r_1(a)=\dfrac{2}{2a+\sqrt{\pi}},
\qquad r_2(a)=\dfrac{4(a+\sqrt{\pi})}{4a^2+7\sqrt{\pi}a+2\pi},
$$
$$
\beta(a)=\dfrac{1}{2}\dfrac{2a+\sqrt{\pi}}{a+\sqrt{\pi}}.
$$

Let us make in the equation (2.3) replacement of  variable
$C=C(\mu),
C'=C(\mu')$, where
$$
C(\mu)=\dfrac{\mu}{1-a|\mu|},\qquad |\mu|<\alpha,\qquad \alpha=\dfrac{1}{a}.
$$

We denote the fuction $h(x,C)$ agian through $h(x,\mu)$.
The equation (2.3) passes in to following equation, standard
for transport equation
$$
\mu\dfrac{\partial h}{\partial x}+h(x,\mu)=\int\limits_{-\alpha}^{\alpha}
\rho(\mu')q(\mu,\mu')h(x,\mu')d\mu',
\eqno{(2.4)}
$$
where
$$
\rho(\mu')=\exp\Big[-\Big(\dfrac{\mu'}{1-a|\mu'|}\Big)^2\Big]
\dfrac{1}{(1-a|\mu'|)^3},
$$
$$
q(\mu,\mu')=r_0(a)+r_1(a)\dfrac{\mu}{1-a|\mu|}\dfrac{\mu'}{1-a|\mu'|}+$$$$+
r_2(a)\Big[\Big(\dfrac{\mu}{1-a|\mu|}\Big)^2-\beta(a)\Big]
\Big[\Big(\dfrac{\mu'}{1-a|\mu'|}\Big)^2-\beta(a)\Big].
$$

Let us notice, that on the ends of  interval of integration
$$
\rho(\pm \alpha)=0,
$$
and, besides,
$$
\lim\limits_{\mu\to\pm \alpha}\rho(\mu)C^n(\mu)=0
$$
for any natural number $n$.

\begin{center}
  \bf 3. Eigen functions and eigen values
\end{center}

Seperation of variables in the equation (2.4), taken in the form
$$
h_\eta(x,\mu)=\exp\Big(-\dfrac{x}{\eta}\Big)\Phi(\eta,\mu), \qquad
\eta \in \mathbb{C},
\eqno{(3.1)}
$$
transforms equation (3.1) to characteristic equation
$$
(\eta-\mu)\Phi(\eta,\mu)=\eta Q(\eta,\mu),\qquad \eta,\mu\in (-\alpha,+\alpha),
\eqno{(3.2)}
$$
where
$$
Q(\eta,\mu)=r_0(a)n_0(\eta)+r_1(a)C(\mu)n_1(\eta)+\hspace{4cm}$$$$+
r_2(a)\Big(C^2(\eta)-\beta(a)\Big)\Big(C^2(\mu)-\beta(a)\Big).
$$

Here
$$
n_\alpha(\eta)=\int\limits_{-\alpha}^{\alpha}\Phi(\eta,\mu)C^\alpha(\mu)
\rho(\mu)d\mu,\qquad \alpha=0,1,2,
\eqno{(3.3)}
$$
is the zero, first and second moments of eigen function with
weight $\rho(\mu)$.

Eigen functions of the continuous spectrum filling
by the continuous fashion an interval $ (-\alpha, \alpha) $,
we find \cite{13} in space of the generalized functions
$$
\Phi(\eta,\mu)=\eta Q(\eta,\mu)P\dfrac{1}{\eta-\mu}+g(\eta)\delta(\eta-\mu),
\quad \eta\in (-\alpha,\alpha).
\eqno{(3.4)}
$$

Here $g(\eta)$ is the unknown function, defined from
equations (3.3), $Px^{-1}$  is the distribution, meaning
principal value of integral
by integration $x^{-1}$, $\delta(x)$ is the Dirac delta-function.

Let us substitute eigen functions (3.4) in normalization equalities
(3.3). We will receive the following system of the dispersion equations
$$
n_\alpha(\eta)+\eta\int\limits_{-\alpha}^{\alpha}Q(\eta,\mu)C^\alpha(\mu)
\rho(\mu)\dfrac{d\mu}{\mu-\eta}=g(\eta)\rho(\eta)C^\alpha(\eta),
\eqno{(3.5)}
$$
$\alpha=0,1,2$.

We denote
$$
t_n(\eta)=\eta \int\limits_{-\alpha}^{\alpha}C^n(\mu)\dfrac{\rho(\mu)d\mu}
{\mu-\eta},\qquad n=0,1,2,3,4.
$$

Now system of the dispersion equations (3.5) it is possible
transform to the form
$$
n_\alpha(\eta)+r_0(a)n_0(\eta)t_\alpha(a)+r_1(a)n_1(\eta)t_1(\eta)+
$$
$$
+(n_2(\eta)-\beta(a)n_0(\eta))(t_{\alpha+2}(\eta)-\beta(a)t_\alpha(\eta))=
g(\eta)\rho(\eta)C^\alpha(\eta),
\eqno{(3.6)}
$$
where $\alpha=0,1,2$.

Let us write down the equations (3.6) in the vector form
$$
\Lambda(\eta)n(\eta)=g(\eta)\rho(\eta)\left[\begin{array}{c}
                                        1 \\
                                        C(\eta) \\
                                        C^2(\eta)
                                      \end{array}\right].
\eqno{(3.7)}
$$

Here $\Lambda(\eta)$ is the dspersion matrix-function with
elements
$$
\lambda_{ij}(\eta)\quad (i,j=1,2,3),
$$
$n(\eta)$ is the normalization vector with elements $n_\alpha(\eta)\quad
(\alpha=0,1,2)$.

Elements of the dispersion matrix in the explicit form will
more low be necessary
$$
\lambda_{11}(z)=1+\Big[r_2(a)+\beta^2(a)r_2(a)\Big]t_0(z)-
\beta(a)r_2(a)t_2(z),
$$
$$
\lambda_{12}(z)=r_1(a)t_1(z),
$$
$$
\lambda_{13}(z)=r_2(a)\Big[-\beta(a)t_0(z)+t_2(z)\Big],
$$

$$
\lambda_{21}(z)=\Big[r_0(a)+\beta^2(a)r_2(a)\Big]t_1(z)-\beta(a)
r_2(a)t_3(z),
$$
$$
\lambda_{22}(z)=1+r_1(a)t_3(z),
$$
$$
\lambda_{23}(z)=r_2(a)\Big[-\beta(a)t_1(z)+t_3(z)\Big],
$$

$$
\lambda_{31}(z)=\Big[r_0(a)+\beta^2(a)r_2(a)\Big]t_2(z)-\beta(a)
r_2(a)t_4(z),
$$
$$
\lambda_{32}(z)=r_1(a)t_3(z),
$$
$$
\lambda_{33}(z)=1+r_2(a)\Big[-\beta(a)t_2(z)+t_4(z)\Big].
$$

We introduce the dispersion function $\lambda(z)$, $\lambda(z)=\det
\Lambda(z)$. In the explicit form we have
$$
\lambda(z)=\lambda_{11}(z)\lambda_{22}(z)\lambda_{33}(z)+
r_1(a)t_3(z)\lambda_{13}(z)\lambda_{21}(z)+
$$

$$
+r_1(a)t_1(z)\lambda_{31}(z)\lambda_{23}(z)-
\lambda_{13}(z)\lambda_{22}(z)\lambda_{31}(z)-
$$

$$
-r_1(a)t_3(z)\lambda_{11}(z)\lambda_{23}(z)-r_1(a)t_1(z)\lambda_{21}(z)
\lambda_{33}(z).
$$

From vector equation (3.7) we find
$$
n_\alpha(\eta)=g(\eta)\rho(\eta)\dfrac{\Lambda_\alpha(\eta)}{\lambda(\eta)},
\quad \alpha=0,1,2,
\eqno{(3.8)}
$$
where $\Lambda_\alpha(\eta)$  is the determinant received from
determinant of system (3.6) by replacement in it $ \alpha $-th
column by the column from free members of this system.
We will write out these
determinants in the explicit form
$$
\Lambda_0(z)=\Lambda_{11}(z)-C(z)\Lambda_{21}(z)+C^2(z)\Lambda_{31}(z)=
\lambda_{22}(z)\lambda_{33}(z)-
$$
$$
-r_1(a)t_3(z)\lambda_{23}(z)-C(z)r_1(a)\Big[t_1(z)\lambda_{33}(z)-
t_2(z)\lambda_{13}(z)\Big]+
$$
$$
+C^2(z)\Big[r_1(a)t_1(z)\lambda_{23}(z)-
\lambda_{22}(z)\lambda_{13}(z)\Big],
$$

$$
\Lambda_1(z)=\Lambda_{12}(z)+C(z)\Lambda_{22}(z)-C^2(z)\Lambda_{32}(z)=
-\lambda_{21}(z)\lambda_{33}(z)+
$$
$$
+\lambda_{31}(z)\lambda_{33}(z)+C(z)\Big[\lambda_{11}(z)\lambda_{33}(z)-
\lambda_{31}(z)\lambda_{13}(z)\Big]-
$$
$$
-C^2(z)\Big[\lambda_{11}(z)\lambda_{23}(z)-
\lambda_{21}(z)\lambda_{13}(z)\Big],
$$

$$
\Lambda_2(z)=\Lambda_{31}(z)-C(z)\Lambda_{32}(z)+C^2(z)\Lambda_{33}(z)=
$$
$$
=r_1(a)t_3(z)\lambda_{21}(z)-\lambda_{31}(z)\lambda_{22}(z)-
C(z)r_1(a)\Big[t_3(z)\lambda_{11}(z)-t_1(z)\lambda_{33}(z)\Big]+
$$
$$
+C^2(z)\Big[\lambda_{11}(z)\lambda_{22}(z)-r_1(a)
t_1(z)\lambda_{21}(z)\Big].
$$

Here $\Lambda_{ij}(z)$ is the minor of element $\lambda_{ij}(z)$.

By means of equalities (3.8) we will transform equality for $Q(\eta, \mu) $
to the form
$$
Q(\eta,\mu)=\tilde Q(\eta,\mu)\dfrac{g(\eta)}{\lambda(\eta)}\rho(\eta),
\eqno{(3.9)}
$$
where
$$
\tilde Q(\eta,\mu) =r_0(a)\Lambda_0(\eta)+r_1(a)C(\mu)\Lambda_1(\eta)+
\hspace{4cm}
$$
$$
+r_2(a)\Big[C^2(\mu)-\beta(a)\Big]\Big[\Lambda_2(\eta)-\beta(a)
\Lambda_0(\eta)\Big].
$$

By means of equality (3.9) we will transform expression (3.4) for
eigen functions
$$
\Phi(\eta,\mu)=\tilde \Phi(\eta,\mu)g(\eta),
\eqno{(3.10)}
$$
where
$$
\tilde \Phi(\eta,\mu)=\eta\dfrac{\tilde Q(\eta,\mu)}{\lambda(\eta)}\rho(\eta)
P\dfrac{1}{\eta-\mu}+\delta(\eta-\mu).
\eqno{(3.11)}
$$

From equality (3.10) it is visible, that eigen functions are defined
accurate within to coefficient -- any function $g (\eta) $,
identically not equal to zero. Owing to uniformity of the initial
kinetic equation it is possible to consider this function identically
equal to unit ($g (\eta) \equiv 1$) and further in quality
eigen function corresponding to  continuous spectrum, it is possible
to consider the functions defined by equality (3.11).
Apparently from the solution of the characteristic equation, continuous
spectrum of the characteristic equation is the set
$$
\sigma_c=\{\eta: -\alpha<\eta<+\alpha\}.
$$

By definition the discrete spectrum of the characteristic
equation consists of set of zero of dispersion function.

Expanding dispersion function in Laurent series in a vicinity
infinitely remote point, we are convinced, that it in this point
has zero of the fourth order. Applying an argument principle \cite{14}
from the theory of functions complex variable, it is possible to show, that
other zero, except $z_i =\infty $, dispersion function not
has. Thus, the discrete spectrum of the characteristic
equations consists of one point $z_i =\infty $,
multiplication factor which it is equal four,
$$
\sigma_d=\{z_i=\infty\}.
$$

To point $z_i =\infty $, as to the 4-fold point of  discrete spectrum,
corresponds following four discrete (partial) solutions
of the kinetic decision (2.4)
$$
h_0(x,\mu)=1,
$$
$$
h_1(x,\mu)=C(\mu),
$$
$$
h_3(x,\mu)=C^2(\mu)-\dfrac{1}{2},
$$
$$
h_3(x,\mu)=(x-\mu)\Big(C^2(\mu)-\dfrac{3}{2}\Big).
$$

Let us result formulas Sokhotsky for the difference and the sum of the boundary
values of dispersion function from above and from below on the
$(-\alpha,+\alpha)$
$$
\lambda^+(\mu)-\lambda^-(\mu)=2\pi i \rho(\mu) \tilde Q(\mu,\mu),\quad
\mu\in (-\alpha,+\alpha),
$$
and
$$
\dfrac{\lambda^+(\mu)+\lambda^-(\mu)}{2}=\lambda(\mu),\quad
\mu\in (-\alpha,+\alpha).
$$

\begin{center}
  \bf 4. The structure of general solusion of kinetic equation
\end{center}

Here we will sum up the done analysis.

We actually prove the theorem of general solution structure of
the equations (2.4).

{\sc Theorem.} The general solution of the equation (2.4)
is the sum of the linear combinations of discrete (partial)
solutions of this equation with
any coefficients and integral on the continuous spectrum
from eigen functions correspond to the continuous spectrum, with
unknown coefficients
$$
h(x,\mu)=A_0h_0(x,\mu)+A_1h_2(x,\mu)+A_2h_2(x,\mu)+A_3h_3(x,\mu)+
$$
$$
+\int\limits_{-\alpha}^{\alpha}\exp\Big(-\dfrac{x}{\eta}\Big)\tilde
\Phi(\eta,\mu)A(\eta)d\eta.
\eqno{(4.1)}
$$

In equality (4.1) $A_\alpha \quad (\alpha=0,1,2,3) $ is the
coefficients correspond to the discrete spectrum, and the unknown
function $A(\eta) $ is the coefficient corresponds
to continuous spectrum.

Coefficients of discrete and continuous spectra are subject
to finding from boundary conditions. In following works authors
assume to solve a number of substantial boundary problems of the kinetic
theories.

Let us consider two partial limiting cases the kinetic
equations.

Let us begin with the case $a=0$. This case corresponds constant
collision frequency  of molecules. In this case essentially becomes simpler
expression for eigen functions of the continuous spectrum
$$
\tilde \Phi (\eta,\mu)=\dfrac{1}{\sqrt{\pi}}\eta\Big(\dfrac{3}{2}-\mu^2\Big)
P \dfrac{1}{\eta-\mu}+e^{\eta^2}\lambda(\eta)\delta(\eta-\mu),
$$
where $\lambda(z)$ is the dispersion function,
expression for which also essentially becomes simpler and has the
following form
$$
\lambda(z)=-\dfrac{1}{2}-(z^2-\dfrac{3}{2})\lambda_C(z),
$$
$\lambda_C(z)$ is the dispersion function of plasma, entered by
Van Kampen in  1955,
$$
\lambda_C(z)=1+\dfrac{z}{\sqrt{\pi}}\int\limits_{-\infty}^{\infty}
\dfrac{e^{-\mu^2}d\mu}{\mu-z}=\dfrac{1}{\sqrt{\pi}}
\int\limits_{-\infty}^{\infty}\dfrac{e^{-\mu^2}\mu d\mu}{\mu-z},
$$
or
$$
\lambda_C(z)=1-2ze^{-z^2}\int\limits_{0}^{z}e^{u^2}du+(-)
\sqrt{\pi}i z^{-z^2},\quad \Im z>0\, (\Im z<0).
$$

In numerical calculations instead of last it is convenient to use
the formula
$$
\lambda_C(z)=1-2z^2\int\limits_{0}^{1}e^{-z^2(1-t^2)}dt+(-)
\sqrt{\pi}i z^{-z^2},\quad \Im z>0\, (\Im z<0).
$$
\begin{figure}[ht]
\begin{flushleft}
\includegraphics[width=16.0cm, height=10cm]{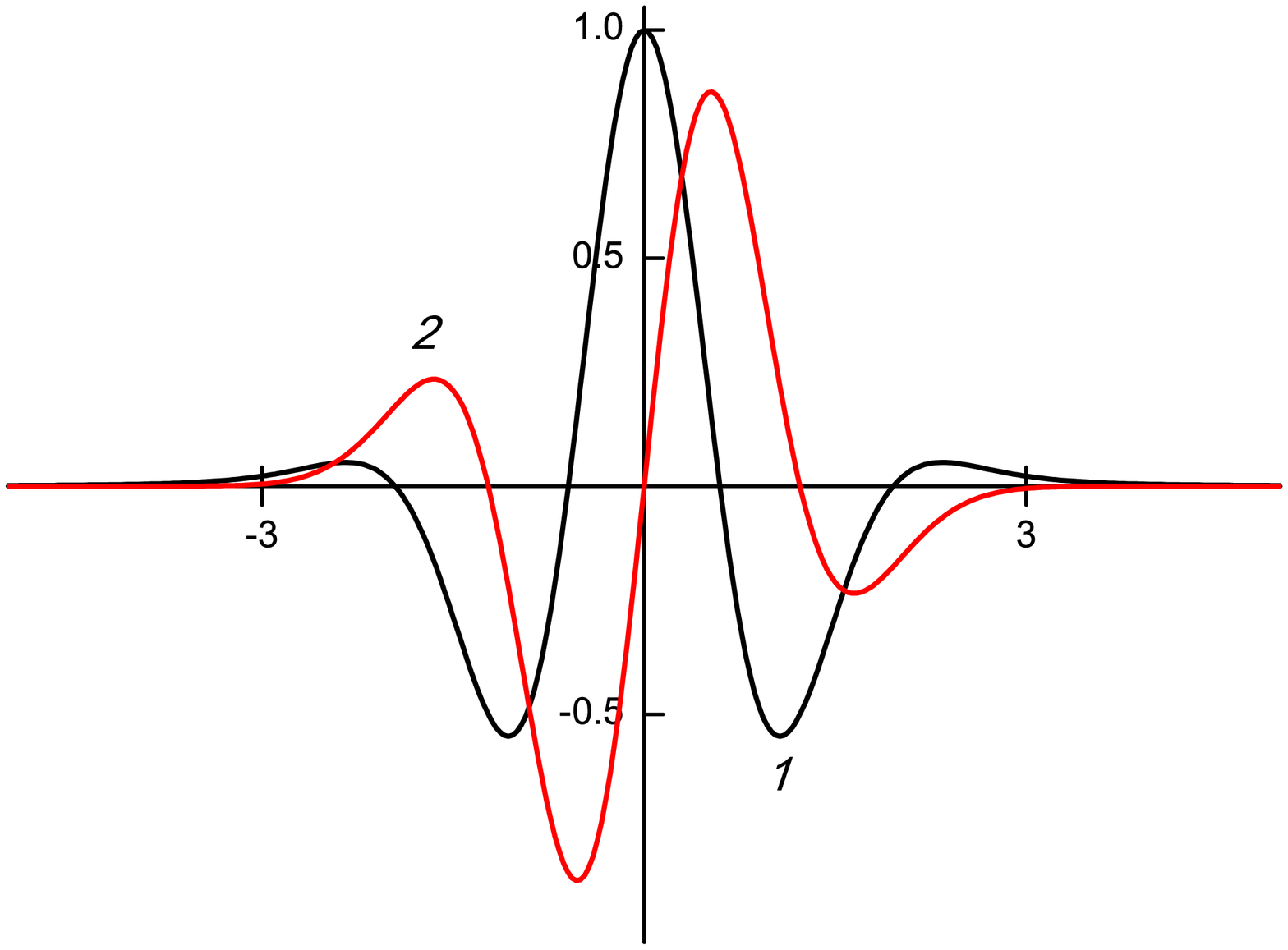}
\end{flushleft}
\center{Fig. The real (curve 1) and imaginare  (curve 2) parts
of boundary values  $\lambda^+(x)$ of dispersion function
on real axis, $a=0$.}
\end{figure}

\clearpage
We investigate also other limiting case, when frequency of collisions
is proportional to the module of molecular velocity, i.e.
$ \nu(C)=\nu_1|C| $. In this case the kinetic equation
passes in the equation
$$
{\rm sign} C \dfrac{\partial h}{\partial x_1}+h(x_1,C)=
\int\limits_{-\infty}^{\infty}e^{-C'^2}|C'|q(C,C')h(x_1,C)dC',
\eqno{(4.2)}
$$
where
$$
q(C,C')=1+CC'+(C^2-1)(C'^2-1).
$$

Further the variable $x_1$ we will again designate through $x $.

Let us search for the solution of the equation (4.2) in the form
$$
h(x,C)=a_1(x)+ \tilde a_1(x){\rm sign} C+
a_2(x)C+ \tilde a_1(x)C{\rm sign} C+$$$$+
a_3(x)(C^2-1)+ \tilde a_3(x)(C^2-1){\rm sign} C.
\eqno{(4.3)}
$$

After substitution (4.3) in the equation (4.2), we receive six
the linear differential equations of the first order
$$
a_1'(x)+\tilde a_1(x)=0,
$$
$$
\tilde a_1'(x)=\dfrac{\sqrt{\pi}}{2}\tilde a_2(x),
$$
$$
a_2'(x)+ a_2(x)=0,
$$
$$
\tilde a_2'(x)=\dfrac{\sqrt{\pi}}{2}\tilde a_1(x)+\sqrt{\pi}a_3(x),
$$
$$
a_3'(x)+ a_3(x)=0,
$$
$$
\tilde a_3'(x)=\dfrac{\pi}{2}\tilde a_2(x).
$$

We solve this system and we find unknown functions $a_j(x)$ and
$\tilde a_j(x)\; (j=1,2,3)$
$$
a_1(x)=-\dfrac{1}{\sqrt{3}\alpha_0}A_0e^{-\alpha_0 x}-\tilde A_1x+
A_1,
$$
$$
\tilde a_1(x)=-\dfrac{1}{\sqrt{3}\alpha_0}A_0e^{-\alpha_0 x}+A_1,
$$
$$
a_2(x)=\dfrac{1}{\alpha_0}A_0e^{-\alpha_0 x}+A_2,
$$
$$
\tilde a_2(x)=A_0e^{-\alpha_0 x},
$$
$$
a_3(x)=-\dfrac{1}{\sqrt{3}\alpha_0}A_0e^{-\alpha_0 x}-\tilde A_3x+
A_3,
$$
$$
\tilde a_3(x)=-\dfrac{1}{\sqrt{3}\alpha_0}A_0e^{-\alpha_0 x}+\tilde A_3.
$$

Here $A_0, A_1, A_2, A_3, \tilde A_1, \tilde A_3$ are arbitrary
constants, $\alpha_0=\dfrac{\sqrt{3\pi}}{2}$.

Thus, the required general solution of the equation (4.2) is constructed and
has the following form
$$
h(x,C)=A_0e^{-\alpha_0 x}\Big[-\dfrac{1}{\sqrt{3}\alpha_0}-
\dfrac{1}{\sqrt{3}}{\rm sign C}+\dfrac{1}{\alpha_0}C+
$$
$$
+C{\rm sign C}-\dfrac{1}{\sqrt{3}\alpha_0}(C^2-1)-\dfrac{1}{\sqrt{3}}
(C^2-1){\rm sign C}\Big]-
$$
$$
-\tilde A_1x+A_1+\tilde A_1 {\rm sign C}+A_2C+
$$
$$
+(-\tilde A_3x+A_3)(C^2-1)+\tilde A_3(C^2-1){\rm sign C}.
$$

\begin{center}
  \bf 4. Conclusion
\end{center}

In the present work the one-dimensional kinetic equation  with
integral of collisions relaxation type BGK (Bhatnagar, Gross and
Krook) is constructed.
Frequency of collisions of molecules as affine depending
on the module molecular velocity is considered.

At construction the equations are used laws of preservation of
number of particles (the numerical density), momentum and energy.
The constructed equation
will be transformed to standard kind of the equation
of type of the equation
carrying over with polynomial kernel.

Separation of variables leads to the characteristic equation.
By means of normalizing equalities the system
of dispersion equations is entered. Its determinant
is called as dispersion function. It is investigated continuous and
discrete spectra of the characteristic equation.

The set of zero of the dispersion equation makes the discrete
spectrum of the characteristic equation. The eigen
solutions of the initial kinetic equation corresponds
to discrete spectrum are found. These solutions so-called discrete
(or partial) solutions.

The solution of the characteristic equation in space of the generalized
functions leads to eigen functions correspond to the continuous
spectrum.

Results of the spent analysis are formulated in the form of the theorem about
structure of the general solution of the entered kinetic equation.

\end{document}